\documentclass[aps,pra,showpacs,twocolumn,superscriptaddress]{revtex4}

\usepackage{graphicx}
\usepackage{dcolumn}
\usepackage{bm}
\usepackage{amsmath}

\begin{document}

\title{Efficient detection of an ultra-bright single-photon source using  superconducting nanowire single-photon detectors }

\author{Rui-Bo Jin}
\email{ruibo@nict.go.jp}
\author{Mikio Fujiwara}
\affiliation{Advanced ICT Research Institute, National Institute of Information and Communications Technology (NICT), 4-2-1 Nukui-Kitamachi, Koganei, Tokyo 184-8795, Japan}

\author{Taro Yamashita }
\author{Shigehito Miki }
\author{Hirotaka Terai }
\affiliation{Advanced ICT Research Institute, National Institute of Information and Communications Technology (NICT), 588-2 Iwaoka, Kobe 651-2492, Japan}
\author{Zhen Wang }
\affiliation{Advanced ICT Research Institute, National Institute of Information and Communications Technology (NICT), 588-2 Iwaoka, Kobe 651-2492, Japan}
\affiliation{Shanghai Institute of Microsystem and Information Technology, Chinese Academy of Sciences (CAS), 865 Changning Road, Shanghai 200050, China}

\author{Kentaro Wakui}
\affiliation{Advanced ICT Research Institute, National Institute of Information and Communications Technology (NICT), 4-2-1 Nukui-Kitamachi, Koganei, Tokyo 184-8795, Japan}

\author{Ryosuke Shimizu}
\affiliation{Center for Frontier Science and Engineering, University of Electro-Communications (UEC), 1-5-1 Chofugaoka, Chofu, Tokyo 182-8585, Japan}

\author{Masahide Sasaki}
\affiliation{Advanced ICT Research Institute, National Institute of Information and Communications Technology (NICT), 4-2-1 Nukui-Kitamachi, Koganei, Tokyo 184-8795, Japan}

\begin{abstract}
We investigate the detection of an ultra-bright single-photon source using highly efficient superconducting nanowire single-photon detectors (SNSPDs) at telecom wavelengths.
Both the single-photon source and the detectors are characterized in detail.
At a pump power of 100 mW (400 mW), the measured coincidence counts can achieve 400 kcps (1.17 Mcps), which is the highest ever reported at telecom wavelengths to the best of our knowledge.
The multi-pair contributions at different pump powers are analyzed in detail.
We compare  the experimental and theoretical second order coherence functions  $g^{(2)}(0)$ and find that the conventional experimentally measured  $g^{(2)}(0)$ values are smaller than the theoretically expected ones.
We also consider the saturation property of SNSPD and find that SNSPD can be easier to  saturate  with a thermal state rather than with a coherent state.
The experimental data and theoretical analysis should be useful for the future experiments to detect ultra-bright down-conversion sources with high-efficiency detectors.
\end{abstract}




\maketitle 

\section{Introduction}

Quantum information and telecommunication technologies  have made vast progress in recent years. For example, the free space teleportation
and entanglement distribution have been demonstrated at 150 km \cite{Yin2012} and multi-photon entangled state has been extended to eight photons \cite{Yao2012, Huang2011}.
However, there are several challenges in these experiments.
One  is that the count rate is very low, so  one needs to accumulate the data for a long time to obtain reliable data.
To overcome such a challenge,  highly efficient single-photon detectors and ultra-bright single-photon sources  need to be developed.

From the perspective of single-photon detectors, many different kinds of detectors have been developed for quantum information applications \cite{Hadfield2009}.
At the near-infrared (NIR) wavelength range (around 800 nm), since the existence of highly efficient silicon avalanche photodiodes (APD, e.g., SPCM, PerkinElmer), many important experiments have been demonstrated at this wavelength range \cite{Pan2012}.
At the telecom wavelength range (around 1550 nm), however, experiments suffered from the low performance of the detectors \cite{Eisaman2011}.
The recent dramatic development of superconducting nanowire single-photon detectors (SNSPDs) \cite{Natarajan2012} with  high system detection efficiency (SDE) can solve this problem.
For example, the NIST group has demonstrated SNSPDs with tungsten silicide (WSi) with a 0.93 SDE at a temperature around 0.12 Kelvin (K) \cite{Marsili2013}.
The MIT group has shown a niobium nitride (NbN) SNSPDs  array  with a  0.76 SDE  at 2.5 K \cite{Rosenberg2013}.
The NICT group has reported SDE of  0.76  in niobium titanium nitride (NbTiN) SNSPDs \cite{Miki2013}, and 0.61-0.80 SDE SNSPDs with low filling-factor at 2.3 K \cite{Yamashita2013}.
Such SNSPDs are promising for quantum information applications because they have the merits of high efficiency, wide spectral response, short recovery time (high speed), low dark count rate, low timing jitter, and free-running operation \cite{Natarajan2012}.

Although the performance of these high-efficiency SNSPDs has been widely characterized in Refs \cite{Marsili2013, Rosenberg2013, Miki2013, Yamashita2013}, in all these reports, SNSPDs were tested and characterized with a classical weak coherent state,  never exploiting their high performance with a quantum light source.
The weak coherent state follows a Poisson distribution, while the single-photon state from a down-conversion source follows a geometric distribution.
These two statics affects to the performance of SNSPDs differently.
Therefore, the first motivation of our experiment is to investigate the performance of such high-efficiency SNSPDs with a single-photon source.
We measure  the spectral range, analyze  the saturation of the SNSPDs with thermal light and coherent light,  and compare  the performance of SNSPDs with commercial APDs.

From the perspective of single-photon sources, the one based on a spontaneous parametric down-conversion (SPDC) process from a periodically poled $\mathrm{KTiOPO_4}$ (PPKTP) crystal has been proved to be very promising in many experiments \cite{Evans2010, Gerrits2011, Eckstein2011, Grice2012, Yabuno2012, Zhong2012, Dixon2013, Jin2013PRA, Jin2013OE, Harder2013, Zhou2013, Jin2013Sagnac}.
The PPKTP crystal has a high nonlinear efficiency and a high damage threshold.
At telecom wavelengths,  the PPKTP crystal satisfies not only the quasi-phase matching condition, but also the group-velocity matching (GVM) condition \cite{Grice1997, Konig2004}.
The single-photon source from a GVM-PPKTP crystal may have a high spectral purity \cite{Evans2010, Gerrits2011, Eckstein2011} and  wide spectral tunability \cite{Jin2013OE}.

In all the previous experiments \cite{Evans2010, Gerrits2011, Eckstein2011, Grice2012, Yabuno2012, Zhong2012, Dixon2013, Jin2013PRA, Jin2013OE, Harder2013, Zhou2013}, however, this highly efficient photon source was detected by low-efficiency or low-speed detectors.
For example, in Refs \cite{Evans2010, Gerrits2011, Eckstein2011, Grice2012, Yabuno2012, Zhong2012, Dixon2013}, the photons were detected by InGaAs APDs with less than or equal to 25\%  quantum efficiency; and
in Refs \cite{Evans2010, Yabuno2012}, to match the low speed of the InGaAs APDs, the repetition rate of the pump laser was decreased from 76 MHz to around 4 MHz, so the performance of this source was not fully demonstrated.
Therefore, the second motivation of this experiment is to fully characterize the performance this highly efficient single-photon source.
We detected the photons with high-efficiency (quantum efficiency of over 70 \%) and high-speed (dead time of around 40 ns) SNSPDs.
We pump the PPKTP crystal with a pump power up to 400 mW,  measure  the second correlation function,  and investigate the multi-photon contribution in this single-photon source.

This paper is organized as follows.
In Section 2, we describe the  detection efficiency, dark count rate and spectral range of our SNSPDs.
In Section 3, the single and coincidence counts of a single-photon source from PPKTP crystal are measured with the SNSPDs.
Then, in Section 4, the multi-pair components in the single-photon source are analyzed.
To investigate the second order coherence function,  $g^{(2)}(0)$, we experimentally measure it in Section 5 and theoretically analyze several different coherence functions in Section 6.
In Section 7, we consider the different responses of SNSPDs with a coherent state and a thermal state.
After that, we compare the performance of SNSPDs and two kinds of commercial InGaAs APDs in Section 8.
Finally, the discussion and conclusion  are in Section 9 and Section 10.

The detailed characterization of the SNSPDs and the GVM-PPKTP source in this paper is of great importance for their future applications.
We believe the combination of high brightness single-photon sources and high performance detectors is the road one must follow in the future development of quantum communication and information  technologies.

\section{Measuring the detecting efficiency  and spectral range of SNSPD}

Our SNSPDs  are fabricated with 5-9 nm thick and 80-100 nm wide NbN or NbTiN  meander nanowire on thermally oxidized silicon substrates \cite{Miki2013, Yamashita2013}.
The nanowire covers an area of 15 $\mu$m  $\times$ 15 $\mu$m. The SNSPDs are installed in a Gifford-McMahon cryocooler system and are cooled to 2.1 K.
The measured timing jitter and  dead time (recovery time) were  68 ps \cite{Miki2013} and 40 ns \cite{Miki2007}.

\begin{figure}[tbp]
\centerline{\includegraphics[width=.9\columnwidth]{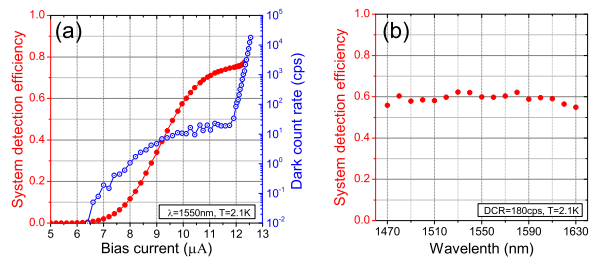}}
\caption{(a) Measured system detection efficiency (SDE) and dark count rate (DCR) as functions of the bias current, with 1550 nm  wavelength at 2.1 K.
(b) Measured system detection efficiency (SDE)  as a function of the wavelength, with dark counts of around 180 cps at 2.1 K.}
\label{SDE}
\end{figure}

We measured the system detection efficiency (SDE, including the fiber coupling efficiency, transmission efficiency and quantum efficiency of the SNSPD chip) and dark count rates (DCR) as a function of the bias currents.
Figure \ref{SDE}(a) shows  typical results measured at 1550 nm, where the SDE can be over 0.60 (0.78) with a dark count rate of around 180 cps (2 kcps).
DCR in the low bias current region in Fig.\,\ref{SDE}(a) is caused by black body radiation at room temperature \cite{Yamashita2010}.
In our previous work  \cite{Miki2013, Yamashita2013}, the DCR was suppressed  to several tens of cps in the high SDE region.

Spectral range is another important parameter for detectors.
Here,  we report our measured results of the SDE at different wavelengths.
Our method is similar to that of Ref \cite{Hadfield2009}.
Wavelength tunable laser (Agilent 81980A  and  Santec ECL-200) with a power of less than 6 dBm was attenuated by about 100 dB by two attenuators (Agilent 81570A).
The attenuated laser is detected by an SNSPD, which is connected to a counter (Tektronix TDS2014).
The SDE is calculated as $SDE = SC/(p\lambda /hc)$, where $SC$ is the single count; $h$ is the Plank constant; $c$ is speed of light; $p$ is the light power; and $\lambda$ is the wavelength.
By changing the central wavelength and repeating the measurement, we can obtain the spectral range of SNSPD from 1470 nm to 1630 nm.
The measured results are shown in  Fig.\,\ref{SDE}(b). The corresponding  dark counts were around 180 cps.
The measured SDEs were between 0.55 to 0.63 for all the wavelengths from 1470 nm to 1630 nm, which was consistent with our previous simulation results \cite{Miki2013}.
This result implies that our SNSPDs have a wide spectral response range that covers  at least the S-, C-, and L-bands in telecom wavelengths.

\section{Measuring single  and coincidence counts}

\begin{figure}[tbp]
\centerline{\includegraphics[width=.95\columnwidth]{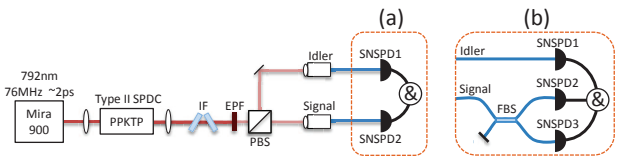}}
\caption{The experimental setup.
Picosecond laser pulses (76 MHz, 792 nm, temporal duration $\sim$2 ps, horizontal polarization) from a mode-locked Titanium sapphire laser (Mira900, Coherent Inc.) were  focused by a $f=200$ mm lens and pumped a  30-mm-long PPKTP crystal with a poling period of 46.1 $\mu$m for type-II group-velocity-matched SPDC.
 PPKTP was maintained at $32.5\,^{\circ}\mathrm{C}$, so as to achieve a degenerate wavelength at 1584 nm.
The down-converted photons, i.e., the signal (H polarization) and idler (V polarization) were collimated by another $f=200$ mm lens, filtered by two interference filters (IF, Thorlabs DMLP1180) and an edge pass filter (EPF, Thorlabs FEL1350),  separated by a polarizing beam splitter (PBS),  and then collected into two single-mode fibers (SMFs).
(a) The signal and idler were  connected to two SNSPDs (SNSPD1 and SNSPD2) and a coincidence counter (Ortec 9353, or Ortec CO4020) for the test of coincidence counts.
(b) The signal was divided by a 50/50 fiber beamsplitter (FBS, Thorlabs 10202A-50-FC) for the test of $g^{(2)}(0)$.
%
%
Since SNSPDs were polarization dependent, the input photons to SNSPD were adjusted by fiber-polarization controllers (not shown).}
\label{setup}
\end{figure}

\begin{figure}[tbp]
\centerline{\includegraphics[width=.95\columnwidth]{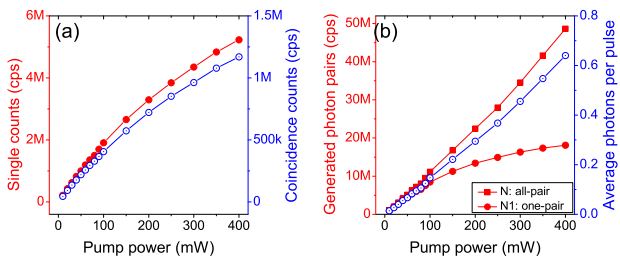}}
\caption{(a) Measured single counts $SC=\sqrt {SC_1  \times SC_2 }$ and coincidence counts as functions of pump power.
(b) Calculated all-pair,  one-pair components generation rates and average photon numbers per pulse as functions of pump power.}
\label{CC}
\end{figure}

Figure\,\ref{setup} shows the experimental setup for the detection of our single-photon source.
To avoid the oscillation of the count rates in high SDE and high DCR regions, we set the bias current at the DCR of less than 1 kcps, and the corresponding SDE of SNSPD1-3 were 0.70, 0.68,  and 0.56 respectively.
First, we measured the single counts and coincidence counts as a function of the pump power, as shown in Fig.\,\ref{CC}(a) and Tab. \ref{tableCC}.
The single counts (coincidence counts) achieved 214 kcps (45 kcps), 1.91 Mcps (406 kcps), and 5.23 Mcps (1.17 Mcps) at pump power of 10 mW, 100 mW and 400 mW, respectively.
As far as we know, these are the highest coincidence counts ever reported at telecom wavelengths.
This count rates were one order higher than our previous results in Ref. \cite{Jin2013PRA}.
For the pump power less than 100 mW, the single counts increased linearly as the pump power increased.
When the pump power was larger than 100 mW, the linearity was slightly degraded, due to not the saturation of SNSPD, but the decrease coupling efficiency, as we will discuss in detail later.
\begingroup
\squeezetable
\begin{table}
\begin{center}
\begin{tabular}{c|ccccccccc}
    \hline
        \hline
 Power & $SC_1$  & $SC_2$   & $CC$     & $\tau $ & $\eta _1$ & $\eta _2$ & $N$ & $N_1$    & $\bar n$ \\
 (mW) & (kcps)   & (kcps)   & (kcps) &         &            &        & (Mcps)& (Mcps)     &   \\
        \hline \hline
 10 & 223 & 205 & 45 & 0.00135 & 0.215 & 0.198 & 1.04 & 1.01 & 0.014 \\
 20 & 447 & 405 & 92 & 0.00132 & 0.219 & 0.198 & 2.05 & 1.95 & 0.027 \\
 30 & 657 & 594 & 136 & 0.00129 & 0.217 & 0.196 & 3.06 & 2.82 & 0.04 \\
 40 & 878 & 772 & 176 & 0.00131 & 0.212 & 0.186 & 4.20 & 3.77 & 0.055 \\
 50 & 1060 & 948 & 219 & 0.00125 & 0.211 & 0.189 & 5.09 & 4.47 & 0.067 \\
 \hline
 60 & 1260 & 1132 & 258 & 0.00127 & 0.204 & 0.183 & 6.27 & 5.35 & 0.083 \\
 70 & 1437 & 1278 & 294 & 0.00124 & 0.203 & 0.180 & 7.21 & 6.02 & 0.095 \\
 80 & 1590 & 1414 & 328 & 0.00120 & 0.202 & 0.179 & 8.04 & 6.57 & 0.106 \\
 90 & 1803 & 1604 & 366 & 0.00124 & 0.194 & 0.172 & 9.53 & 7.52 & 0.125 \\
 100 & 2025 & 1800 & 406 & 0.00128 & 0.187 & 0.165 & 11.15 & 8.48 & 0.147 \\
 \hline
 150 & 2818 & 2500 & 573 & 0.00121 & 0.174 & 0.154 & 16.79 & 11.26 & 0.221 \\
 200 & 3510 & 3093 & 721 & 0.00114 & 0.164 & 0.144 & 22.44 & 13.38 & 0.295 \\
 250 & 4109 & 3591 & 851 & 0.00108 & 0.155 & 0.135 & 27.96 & 14.94 & 0.368 \\
 300 & 4655 & 4060 & 963 & 0.00104 & 0.143 & 0.124 & 34.56 & 16.33 & 0.455 \\
 350 & 5189 & 4502 & 1078 & 0.00101 & 0.134 & 0.115 & 41.59 & 17.37 & 0.547 \\
 400 & 5626 & 4865 & 1170 & 0.00098 & 0.125 & 0.107 & 48.62 & 18.08 & 0.640 \\
  \hline
   \hline
\end{tabular}
\caption{\label{tableCC} Parameters at different pump powers. }
\end{center}
\end{table}
\endgroup

In a lower pump region, we can ignore the multi-pair emission in SPDC and simply estimate the generated photon pairs  with the following equation \cite{Tanzilli2001}: $N = \frac{{SC_1  \times SC_2 }}{{CC_{12}}}$,
where $SC_1$ and $SC_2$ are the single counts from SNSPD1 and SNSPD2, respectively, and $CC_{12}$ is their coincidence counts.
However, we must consider the multi-pair contribution in SPDC, which is especially important for  the high pump power region.
Next, we consider the general case and calculate  the multi-pair components in detail.

\section{Analysis of multi-pair contribution in SPDC}

The output state of the SPDC can be expressed on the basis of photon number as \cite{Kok2007, Broome2011}
\begin{equation}
\label{eq1}
\begin{array}{l}
\left| {\psi _{SPDC} } \right\rangle  = \sqrt {1 - \gamma ^{2} } \sum\limits_{n=0}^\infty  {\gamma ^{n} } \left| {n,n} \right\rangle  \\
 = \sqrt {1 - \gamma ^2 } (\left| {0,0} \right\rangle  + \gamma \left| {1,1} \right\rangle  + \gamma ^2 \left| {2,2} \right\rangle  + \gamma ^3 \left| {3,3} \right\rangle  + ...),  \\
\end{array}
\end{equation}
where $|nn\rangle\equiv|n\rangle_s\otimes|n\rangle_i$ denotes the $n$-pair state containing $n$ photons in both signal and idler modes.
The parameter $\gamma$ is defined as
\begin{equation}\label{eq2}
\gamma  = \sqrt {p\tau },
\end{equation}
 where $p$ is the pump power, and $\tau$ is a constant which is determined by the interaction in the nonlinear medium \cite{Broome2011}.
The probability for the $n$-pair photons per pulse is
\begin{equation}\label{eq3}
Pr(n) = (1 - \gamma ^2 )\gamma ^{{2n}}  = (1 - p\tau )(p\tau )^{n}.
\end{equation}
For bucket detectors, i.e., photon-number-non-resolving detectors, the single counts (SC) and coincidence counts (CC) can be calculated as
\begin{equation}\label{eq4}
 SC_1 = f \times \sum\limits_{n = 1}^\infty  ( 1 - (1 - \eta_1 )^n )(1 - p\tau )(p\tau )^{\rm{n}},
\end{equation}
\begin{equation}\label{eq5}
 SC_2 = f \times \sum\limits_{n = 1}^\infty  ( 1 - (1 - \eta_2 )^n )(1 - p\tau )(p\tau )^{\rm{n}},
\end{equation}
 and
\begin{equation}\label{eq6}
 CC = f \times \sum\limits_{n = 1}^\infty  ( 1 - (1 - \eta_1 )^n ) ( 1 - (1 - \eta_2 )^n )(1 - p\tau )(p\tau )^{\rm{n}},
\end{equation}
where, $f = 76$ MHz is the repetition rate of the laser, $\eta_{1(2)}$ is the overall efficiency, which includes the SDE, transmission efficiency and coupling efficiency \cite{Kok2007, Broome2011} .
We can calculate the exact values of $\eta$ and $\tau$  with Eqs.(\ref{eq4}-\ref{eq6}) for different pump powers.
Then,  all-pair  generation rates can be calculated as
\begin{equation}\label{eq7}
N = f \times \sum_{n = 1}^\infty  {nPr(n) = } f \times \sum_{n = 1}^\infty  n (1 - p\tau )(p\tau )^{\rm{n}}.
\end{equation}
The one-pair  generation rates can be obtained as
\begin{equation}\label{eq8}
N_1  = f \times Pr(1) = f \times (1 - p\tau )(p\tau ).
\end{equation}
The average photon per pulse is calculated as
\begin{equation}\label{eq9}
\bar n = \sum_{n = 1}^\infty nPr(n) = \sum_{n = 1}^\infty n (1 - p\tau )(p\tau )^{n}.
\end{equation}
The calculated results of $\tau$ and $\eta_{1(2)}$ at different pump powers are shown in  Tab. \ref{tableCC}.
Table \ref{tableCC} shows that, the parameters $\eta_{1(2)}$ and $\tau$ decrease as  pump powers increase.
This might be caused by the phenomenon of gain-induced diffraction (GID) \cite{Anderson1995}.
In the case of SPDC in PPKTP bulk crystal, the pump laser is a Gaussian beam.
At a higher pump power, the center portion of the down-converted photons experiences higher gains than the wings, thus distorting the spatial profile and lowering the coupling efficiency \cite{Anderson1995}.
Therefore,  the coupling efficiency needs to be optimized at higher pump powers.
We will experimentally investigate the phenomenon of GID in  future work.

\begin{figure}[tbp]
\centerline{\includegraphics[width=.95\columnwidth]{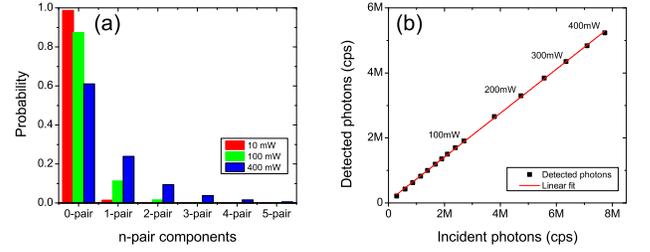}}
\caption{(a) The probabilities of the $n$-pair components  at pump power of 10 mW, 100 mW, and 400 mW.
(b) The detected photon numbers ($\sqrt {SC_1  \times SC_2 }$) as a function of the calculated incident photons ($ N \sqrt {\eta_1\eta_2 /SDE_1/SDE_2} $).}
\label{pandn}
\end{figure}

We also compare the probabilities of zero-pair to five-pair components at pump power of 10 mW, 100 mW, and 400 mW in  Fig.\,\ref{pandn}(a),
where the probabilities of multi-pair components increase rapidly  as pump power increases.
These multi-pair components at high pump power are useful  for multi-photon entangled state generation \cite{Radmark2009} or multi-photon interferometry \cite{Yabuno2012} from a single SPDC source at telecom wavelengths.

\section{Measuring second order correlation function $g^{(2)}(0)$}

\begin{figure}[tbp]
\centerline{\includegraphics[width=.95\columnwidth]{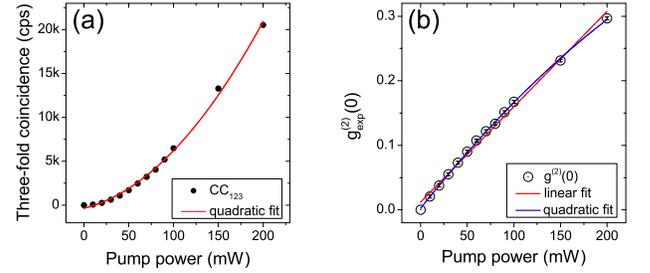}}
\caption{(a) Measured three-fold  coincidence counts $CC_{123}$ as a function of pump power.  (b) Measured $g^{(2)}_{exp}(0)$  as a function of pump power. The uncertainties  were derived using Poissonian errors on the single and coincidence counts.}
\label{g2}
\end{figure}

We also measured the second order coherence function $g^{(2)}(0)$, which is a parameter to characterize the multi-photon components in the photon source \cite{URen2005a}.
The setup is shown in Fig.\,\ref{setup}(b).
 $g^{(2)}(0)$ can be measured  with the below equation \cite{URen2005a, MosleyPhD}:
\begin{equation}\label{eq10}
   g_{exp}^{(2)}(0)  = \frac{{\left\langle {\hat a^\dag  \hat a^\dag  \hat a\hat a} \right\rangle }}{{\left\langle {\hat a^\dag  \hat a} \right\rangle ^2 }} \approx \frac{{2CC_{123}  \times SC_1 }}{{(CC_{12}  + CC_{13} )^2}},
\end{equation}
where  $SC_1$ is the single count of SNSPD1; $CC_{12}$ and $CC_{13}$ are the coincidence counts of SNSPD1 and SNSPD2, and SNSPD1 and SNSPD3, respectively;  and $CC_{123}$ is the three-fold  coincidence counts of SNSPD1, SNSPD2, and SNSPD3.
As shown in Fig.\,\ref{g2}(a), the measured $CC_{123}$ were 77 cps, 6.5 kcps and 20.5 kcps at the pump power of 10 mW, 100 mW, and 200 mW, respectively.
Figure \,\ref{g2}(a)  clearly shows  a quadratic function shape, because the two-pair components in the output state of SPDC are proportional to the square of the pump power \cite{Krischek2010}.
Figure \,\ref{g2}(b) and Tab. \ref{tableCC} show the value of  $g^{(2)}(0)$.
The values were 0.02, 0.17, and  0.30 at  the pump power of 10 mW, 100 mW, and 200 mW, respectively.
Interestingly,  the $g^{(2)}(0)$ data shows a near-linear function shape, different from the quadratic function in Fig.\,\ref{g2}(a).
This can be roughly understood with the following theoretical consideration: at a low pump power,
$ g^{(2)}(0)  = \frac{{\left\langle {\hat a^\dag  \hat a^\dag  \hat a\hat a} \right\rangle }}{{\left\langle {\hat a^\dag  \hat a} \right\rangle ^2 }}  \propto \frac{{Pr_2 }}{{Pr_1^2 }}, $
where $Pr_1$ and $Pr_2$  are the one-pair and two-pair components in the output state of SPDC, respectively.
Both the numerator and denominator are proportional to the square of the pump power.
Therefore, $g^{(2)}(0)$ is a near-linear function of the pump power.
Next we will investigate the accurate relationship between  $g_{exp}^{(2)}(0)$ and pump power.
\begingroup
\squeezetable
\begin{table}[tbp]
\begin{center}
\begin{tabular}{c|ccccccccc}
    \hline
        \hline
 Power & $g^{(2)}_{exp}(0)$ & $g^{(2)}_{exp2}(0)$  &  $g^{(2)}_{s_h}(0)$  & $g^{(2)}_{s}(0)$    &  $g^{(2)}_{s,i}(0)$  & $g^{(3)}_{s,i}(0)$  & $g^{(3)}_{s}(0)$ \\
 (mW) &   heralded      & heralded    & heralded      & thermal     &  squeezed     & squeezed     &   thermal\\
        \hline \hline
 10 & 0.021 & 0.023 & 0.027 & 2.0 & 38.593 & 225.56 & 6.0 \\
 20 & 0.038 & 0.045 & 0.053 & 2.0 & 20.501 & 117.01 & 6.0 \\
 30 & 0.055 & 0.065 & 0.077 & 2.0 & 14.432 & 80.59 & 6.0 \\
 40 & 0.073 & 0.086 & 0.105 & 2.0 & 11.052 & 60.31 & 6.0 \\
 50 & 0.090 & 0.102 & 0.125 & 2.0 & 9.469 & 50.82 & 6.0 \\
   \hline
 60 & 0.108 & 0.122 & 0.152 & 2.0 & 8.058 & 42.35 & 6.0 \\
 70 & 0.122 & 0.137 & 0.173 & 2.0 & 7.267 & 37.60 & 6.0 \\
 80 & 0.133 & 0.150 & 0.191 & 2.0 & 6.728 & 34.37 & 6.0 \\
 90 & 0.151 & 0.172 & 0.223 & 2.0 & 5.989 & 29.93 & 6.0 \\
 100 & 0.167 & 0.194 & 0.256 & 2.0 & 5.408 & 26.45 & 6.0 \\
  \hline
 150 & 0.231 & 0.259 & 0.362 & 2.0 & 4.263 & 19.58 & 6.0 \\
 200 & 0.297 & 0.310 & 0.456 & 2.0 & 3.693 & 16.16 & 6.0 \\
 250 & - & 0.351 & 0.538 & 2.0 & 3.359 & 14.15 & 6.0 \\
 300 & - & 0.391 & 0.625 & 2.0 & 3.099 & 12.60 & 6.0 \\
 350 & - & 0.425 & 0.707 & 2.0 & 2.914 & 11.48 & 6.0 \\
 400 & - & 0.452 & 0.780 & 2.0 & 2.782 & 10.69 & 6.0 \\
  \hline
   \hline
\end{tabular}
\caption{\label{tableg2} Different kinds of second order correlation functions results and third order correlation functions. }
\end{center}
\end{table}
\endgroup

\section{Analysis of theoretical $g^{(2)}(0)$ and $g^{(3)}(0)$ }

Second and third order correlation functions are important parameters for characterizing a photon source.
Besides the experimentally measured $g^{(2)}_{exp}(0)$ in Eq.(\ref{eq4}), we give several theoretical   equations of $g^{(2)}(0)$ and $g^{(3)}(0)$ based on different states, as shown below.
%
%
%
%
%
\begin{itemize}
  \item $g_{exp2}^{(2)} (0) $ is in the form of Eq.(\ref{A-5-7}), which is deduced with the parameters obtained in coincidence counts.
  \item $g^{(2)} _{s_h } (0) = \frac{{\sum_{n = 1}^\infty  {n(n - 1)} Pr_{n - 1} }}{{(\sum_{n = 1}^\infty  n Pr_{n - 1} )^2 }}$, for the heralded signal.  See Eq.(\ref{A-3-4}).
  \item $g^{(2)}_s (0) = \frac{{\sum_{n = 0}^\infty  {n(n - 1)} Pr_n }}{{(\sum_{n = 0}^\infty  n Pr_n )^2 }}$, for the unheralded signal. See Eq.(\ref{A-2-2}).
  \item $g^{(2)}_{s,i} (0) = \frac{{\sum_{n = 0}^\infty  {(2n)(2n - 1)} Pr_n }}{{(\sum_{n = 0}^\infty  {2n} Pr_n )^2 }}$, for the signal and idler. See Eq.(\ref{A-1-3}).
  \item $g_{s,i}^{(3)} (0) = \frac{{\sum\limits_{n = 0}^\infty  {((2n)^3  - 3(2n)^2  + 2(2n))} Pr_n }}{{(\sum\limits_{n = 0}^\infty  {(2n)} Pr_n )^3 }} $, for the signal and idler. See Eq.(\ref{A-6-2}).
  \item $g_s^{(3)} (0) = \frac{{\sum\limits_{n = 0}^\infty  {(n^3  - 3n^2  + 2n)} Pr_n }}{{(\sum\limits_{n = 0}^\infty  n Pr_n )^3 }} $, for the unheralded signal.  See Eq.(\ref{A-6-3}).
\end{itemize}
All these equations are based on the parameters of $p$, $\tau $  and $\eta_{1(2)}$  in Tab. \ref{tableCC}. See more details in Appendix.
The calculated results are listed in Tab. \ref{tableg2}.

Also notice that the $g^{(2)}_{exp2}(0)$  values basically equal  the  $g^{(2)}_{exp}(0)$  values, suggesting the measured data in  Fig.\,\ref{setup} (a) and Fig.\,\ref{setup} (b) are consistent.
However, theoretical values of $g^{(2)}_{s_h}(0)$ are always much bigger than $g^{(2)}_{exp}(0)$, suggesting that the conventional measurement method in  Eq.(\ref{eq10}) \cite{URen2005a, MosleyPhD} is not accurate, especially for high pump power.
%
%

\section{Saturation of SNSPDs }

In  Fig.\,\ref{CC}(a), the linearity of the single counts  decreases dramatically when the pump power increases.
The SNSPDs seem to saturate at a higher pump power.
However, when we change the x-axis to be incident photon numbers, as shown in Fig.\,\ref{pandn}(b), the linearity is almost perfectly maintained at a higher pump power.
In the calculation for Fig.\,\ref{pandn}(b), we assume the SDE is constant, because the measured recovery time of our SNSPDs were 40 ns \cite{Miki2007}, which corresponds to a saturation rate of 25 MHz, much higher than the 5.6 MHz maximum single counts in Tab. \ref{tableCC}.

Furthermore, we consider the saturation of SNSPDs with a coherent state and a thermal state.
In the conventional test of SNSPDs, as shown in previous research \cite{Marsili2013, Rosenberg2013, Miki2013, Yamashita2013}, the  performance of SNSPDs was evaluated with a weak coherent state.
However a coherent state has different statics from a thermal state, which is the case of the signal (idler) photons in SPDC \cite{Jin2013PRA2}.
Therefore, it is meaningful to compare the performance, especially the saturation property, of SNSPDs with these two different states.
%


We rewrite the output state of SPDC in  Eq.(\ref{eq1}) as
\begin{equation}\label{A-7-1}
\left| \varphi  \right\rangle= a_0\left| {00} \right\rangle+ a_1 \left| {11} \right\rangle+ a_2 \left| {22}\right\rangle+ ...\, ,
\end{equation}
The photon number probability follows the geometric distribution given by $ \left| a_n \right|^2= (1 - \gamma ^2 )\gamma ^{2n}  = \frac{1}{{1 + \mu }}(\frac{\mu }{{1 + \mu }})^n$, where $\mu  = \gamma ^2 /(1 - \gamma ^2 )$ is the mean photon number in the signal (or idler) mode.
When this thermal state is detected by SNSPD with an efficiency of $\eta$, the single count is
\begin{equation}\label{A-7-2}
SC_{thermal} = f \times \sum\limits_{n = 0}^\infty  {(1 - (1 - \eta )^n ) \times n}  \times \frac{1}{{1 + \mu }}(\frac{\mu }{{1 + \mu }})^n
\end{equation}

The weak coherent state of LO can be written as
\begin{equation}\label{A-7-3}
\left| \alpha  \right\rangle= c_0 \left| 0 \right\rangle+ c_1 \left| 1 \right\rangle+ c_2 \left| 2 \right\rangle+ ...\, ,
\end{equation}
where  $|n\rangle$ represents the $n$-photon state in the LO mode.
The photon number probability follows the Poisson distribution given by $\left| c_n  \right|^2 = e^{-\nu} \nu^n / n!$, where $\nu$ is the mean photon number in the LO mode.
When this thermal state is detected by SNSPD with an efficiency of $\eta$, the single count is
\begin{equation}\label{A-7-4}
SC_{coherent} = f \times \sum\limits_{n = 0}^\infty  {(1 - (1 - \eta )^n ) \times n}  \times e^{ - \nu } \nu ^n /n!
 \end{equation}

\begin{figure}[tbp]
\centerline{\includegraphics[width=.5\columnwidth]{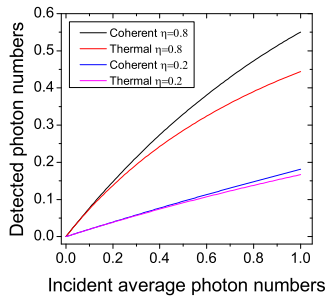}}
\caption{Detected photons as a function of incident photons for coherent state and thermal state with different detection efficiency $\eta$.}
\label{thermalVcoherent}
\end{figure}

With Eq.(\ref{A-7-2}) and Eq.(\ref{A-7-4}), we draw the  detected photons versus incident photons for these two states in Fig.\,\ref{thermalVcoherent}.
Here, we assume both the coherent  and thermal photon sources are pulsed and have  repetition rates  less than the inverse of the recovery time of the SNSPDs.
Both sources have the same average photons per pulse.
Figure\,\ref{thermalVcoherent} shows that as the  average photons per pulse increases, the thermal state saturates faster than the coherent state.
The higher the efficiency $\eta$, the  larger the  difference  between these two states.
This phenomenon can also be understood in the following way: the coherence time of the thermal state (bunched) is shorter than the coherent state (not bunched), so the  detectors  are easier to saturate.
This means we should pay attention to the different response properties of SNSPDs for states with different photon statics.
This phenomenon should be useful for many applications, e.g., the modeling of a practical quantum key distribution system.

\section{Comparison with InGaAs APD}

To show the performance of our SNSPDs, we also compare them with the commercially available InGaAs APD detectors  ID220 and ID210 from ID Quantique.
%
We changed the two SNSPDs in Fig.\,\ref{setup}(a) to two ID220s and two ID210s, and repeated the detection of coincidence counts.
The results are listed in Tab. \ref{table1}, with a pump power of 10 mW.
We confirmed that the coincidence counts of SNSPDs were 30 times higher than those of the InGaAs APDs.

\begin{table}[tbp]
\begin{center}
\begin{tabular}{l|lll}
\hline \hline
   parameters  & SNSPD & ID220 & ID210 \\
 \hline \hline
  SC            & 214 kcps        &  65 kcps                      &  60 kcps \\
  CC            & 45 kcps         &  1.5 kcps                     &  1.1 kcps \\
  Mode          & free-running    &  free-running                 &  external gating\\
  Temperature   &  2.1 K          & $-50\,^{\circ}\mathrm{C}$     &  $-50\,^{\circ}\mathrm{C}$  \\
  Dead time     &  40 ns          &  5 $\mu$s                     &  10 $\mu$s\\
  Dark counts   &  $\sim$250 cps  & $\sim$2 kcps                  & $\sim$2 kcps \\
  Efficiency    & 0.65@1584 nm    &  0.2@1550 nm                  &  0.25@1550 nm\\

 \hline \hline
\end{tabular}
\caption{\label{table1} Comparison of three kinds of detectors using our photon source with a pump power of 10 mW.   $SC=\sqrt {SC_1  \times SC_2 }$.  ID210s were gated by 76 MHz synchronized electronic signal from Mira900, with a gate width of 2 ns.   }
\end{center}
\end{table}

\section{Discussion and Outlook}

Compared with the previous PPKTP source at telecom wavelengths in the work of Evans et al \cite{Evans2010}, the calculated generation rates of photon pairs per second are comparable, but our detected coincidence counts are about 90 times higher.
It is also noteworthy to compare the brightness of our source with the previous results at NIR wavelengths.
In the free-space quantum teleportation experiment by Yin et al \cite{Yin2012}, the coincidence was 440 kcps (after bandpass filtering) with ultra-violet (UV) pump power of 1.3 W.
In the eight-photon entangled state experiment by Huang et al \cite{Huang2011}, the coincidence of 220 kcps (after bandpass filtering) with UV pump power of 300 mW.
In another eight-photon entangled state experiment by Yao \cite{Yao2012}, the coincidence was 1 Mcps (without filtering) with an UV pump power of 1 W.
We noticed that the maximum coincidence counts in our system are comparable to the results of the above experiments \cite{Yin2012, Huang2011, Yao2012}.
This means using SNSPDs and PPKTP crystal at telecom wavelengths  can make the multi-photon coincidence counts similar to or even higher than  those in the NIR range.

We noticed that a combination of gated InGaAs APDs and PPKTP waveguide can also achieve a high coincidence counts at a low pump power \cite{Zhong2012}.
Such a gated InGaAs APD has a relatively low quantum efficiency (25\%) and a high gating rate (625MHz).

Although the  0.93 SDE of in Marsili et al \cite{Marsili2013} is higher than those of others\cite{Rosenberg2013, Miki2013, Yamashita2013}, the sub-Kelvin operating temperature is needed for its best performance.
%
Our  SNSPD \cite{Miki2013, Yamashita2013} operating in a Gifford-McMahon cryocooler at 2.1 K with reasonably high SDE (0.61 - 0.80)  can have a  simpler operation and lower cost for cooling systems, offering a good option for wider applications.
The combination of this highly efficient SNSPD and the highly bright photon source in our scheme  opens the door for  various  future applications at telecom wavelengths, e.g., multi-photon interference in fiber networks and eye-safe free-space quantum key distribution.

\section{Conclusion}
In summary, we have characterized the performance highly efficient SNSPDs and an ultra-bright single-photon source from PPKTP crystal.
The  measured spectral response of the SNSPD can cover 1470 nm to 1630 nm.
The coincidence counts  achieved 0.4 Mcps (1.17 Mcps) at a pump power of 100 mW (400 mW).
The multi-pair emissions in SPDC were analyzed in detail.
We also compared several different  second  and third order coherence functions  at different pump powers.
It was found that $g^{(2)}(0)$ measured from the conventional equation  was smaller than the theoretically expected one.
We considered the different responses of SNSPD with a coherent state and a thermal state, and found that the thermal state saturates faster than the coherent state.
Furthermore, we compared the SNSPDs with the commercial InGaAs APDs and found the coincidence was 30 times higher.
The combination of GVM-PPKTP crystal and SNSPD is important for future quantum information and communication applications at telecom wavelengths.

\section*{Acknowledgements}
R.-B. Jin thanks Y. Shikano for helpful discussions.
This work was supported by the Founding Program for World-Leading Innovative R\&D on Science and Technology (FIRST).

\appendix
\section{ $g^{(2)}(0)$ for both the signal and idler (squeezed state): $g^{(2)}_{s,i} (0)$ }
The definition of $g^{(2)}(0)$ is
\begin{equation}\label{A-1-1}
g^{(2)} (0) \equiv \frac{{\left\langle {\hat a^\dag  \hat a^\dag  \hat a\hat a} \right\rangle }}{{\left\langle {\hat a^\dag  \hat a} \right\rangle ^2 }}{\rm{ = }}\frac{{\left\langle {\hat a^\dag  \hat a\hat a^\dag  \hat a} \right\rangle  - \left\langle {\hat a^\dag  \hat a} \right\rangle }}{{\left\langle {\hat a^\dag  \hat a} \right\rangle ^2 }} = \frac{{\left\langle {\hat n(\hat n - 1)} \right\rangle }}{{\left\langle {\hat n} \right\rangle ^2 }}
\end{equation}
where $\hat a^\dag$ and $\hat a$ are the photon creation and annihilation operators; $\hat n$  is the photon number operator.
The output state of SPDC can be expressed on the number state base as,
\begin{equation}\label{A-1-2}
\begin{array}{l}
\left| {\psi _{SPDC} } \right\rangle  = \sqrt {1 - \gamma ^2 } (\left| {0,0} \right\rangle  + \gamma \left| {1,1} \right\rangle  + \gamma ^2 \left| {2,2} \right\rangle  + ...)\\
= \sqrt {1 - \gamma ^2 } \sum_{n = 0}^\infty  {\gamma ^n } \left| {n,n} \right\rangle  = \sum_{n = 0}^\infty  {\sqrt {Pr_n } } \left| {2n} \right\rangle,\\
 \end{array}
\end{equation}
This is also a typical squeezed state, in which the photon numbers are squeezed.
It can be directly generated by using type I SPDC.
In type II SPDC, this squeezed state can be prepared by combining the signal and idler photons, as shown by Gerrits et al \cite{Gerrits2011}.
We can calculate
$
\left\langle {\hat n} \right\rangle  = \left\langle {\psi _{SPDC} } \right|\hat n\left| {\psi _{SPDC} } \right\rangle  = \sum\limits_{n = 0}^\infty  {2nPr_n }  = \sum\limits_{n = 0}^\infty  {2n} (1 - \gamma ^2 )\gamma ^{{2n}}
$
and
$
\left\langle {\hat n^2 } \right\rangle  = \left\langle {\psi _{SPDC} } \right|\hat n^2 \left| {\psi _{SPDC} } \right\rangle  = \sum\limits_{n = 0}^\infty  {(2n)^2 Pr_n }  = \sum\limits_{n = 0}^\infty  {(2n)^2 } (1 - \gamma ^2 )\gamma ^{{2n}}
$.
Therefore, the $g^{(2)}(0)$ can be obtained as
\begin{equation}\label{A-1-3}
g^{(2)}_{s,i} (0) = \frac{{\left\langle {\hat n(\hat n - 1)} \right\rangle }}{{\left\langle {\hat n} \right\rangle ^2 }} = \frac{{\sum_{n = 0}^\infty  {(2n)(2n - 1)} Pr_n }}{{(\sum_{n = 0}^\infty  {2n} Pr_n )^2 }}.
\end{equation}
At a low pump power and only consider the first order term (one-pair contribution), $g^{(2)}_{s,i} (0)$ can be near infinity $\infty$.
\begin{equation}\label{A-1-4}
g^{(2)}_{s,i} (0)  \to \frac{{2Pr_1 }}{{(2Pr_1 )^2 }}{\rm{ = }}\frac{1}{{2Pr_1 }}{\rm{ = }}\frac{1}{{2(1 - \gamma ^2 )\gamma ^2 }} \to \infty.
\end{equation}

\section{ $g^{(2)}(0)$ for the unheralded  signal (thermal state): $g^{(2)}_s (0) $ }
If we only consider the signal, which is in the subspace of $\left| {\psi _{SPDC} } \right\rangle $,
the density matrix of signal can be written as
\begin{equation}\label{A-2-1}
\begin{array}{l}
\rho _s  = Tr_i (\left| {\psi _{SPDC} } \right\rangle \left\langle {\psi _{SPDC} } \right|) = \sum_{n = 0}^\infty  {Pr_n } \left| n \right\rangle _s \left\langle n \right| \\
= \sum_{n = 0}^\infty  ( 1 - \gamma ^2 )\gamma ^{2n} \left| n \right\rangle _s \left\langle n \right|.\\
\end{array}
\end{equation}
We can calculate  $ \left\langle {\hat n} \right\rangle  = Tr(\hat n\rho _s ) = \sum_{n = 0}^\infty  n Pr_n  = \sum_{n = 0}^\infty  n (1 - \gamma ^2 )\gamma ^{{\rm{2n}}} $
and $\left\langle {\hat n^2 } \right\rangle  = Tr(\hat n^2 \rho _s ) = \sum_{n = 0}^\infty  {n^2 } Pr_n  = \sum_{n = 0}^\infty  {n^2 } (1 - \gamma ^2 )\gamma ^{{\rm{2n}}} $.
Therefore, the $g^{(2)}(0)$ can be obtained as
\begin{equation}\label{A-2-2}
g^{(2)}_s (0) = \frac{{\sum_{n = 0}^\infty  {n(n - 1)} Pr_n }}{{(\sum_{n = 0}^\infty  n Pr_n )^2 }}.
\end{equation}
At low pump power, $g^{(2)}_s (0)$ is near 2.
\begin{equation}\label{A-2-3}
g^{(2)}_s (0) \to \frac{{2Pr_2 }}{{(Pr_1 )^2 }} = \frac{{2(1 - \gamma ^2 )\gamma ^4 }}{{((1 - \gamma ^2 )\gamma ^2 )^2 }} = \frac{2}{{(1 - \gamma ^2 )}} \to 2.
\end{equation}

\section{ $g^{(2)}(0)$ for the heralded signal (heralded single-photon state): $g^{(2)} _{s_h } (0)$}
When at least one photon is detected in the idler channel, the zero-pair component in $\left| {\psi _{SPDC} } \right\rangle $ is removed
\begin{equation}\label{A-3-1}
\left| {\psi _{SPDC} } \right\rangle  \to \sqrt {1 - \gamma ^2 } (\gamma \left| {1,1} \right\rangle  + \gamma ^2 \left| {2,2} \right\rangle  + \gamma ^3 \left| {3,3} \right\rangle  + ...).
\end{equation}
After the coefficients have been normalized, the state can be written as
\begin{equation}\label{A-3-2}
\begin{array}{l}
\left| {\psi _{SPDC_h} } \right\rangle =\sqrt {1 - \gamma ^2 } (\left| {1,1} \right\rangle  + \gamma \left| {2,2} \right\rangle  + \gamma ^2 \left| {3,3} \right\rangle  + ...) \\
 = \sqrt {1 - \gamma ^2 } \sum_{n = 1}^\infty  {\gamma ^{{\rm{n - 1}}} } \left| {n,n} \right\rangle  = \sum_{n = 1}^\infty  {\sqrt {Pr_{n - 1} } } \left| {n,n} \right\rangle.\\
 \end{array}
 \end{equation}
The density matrix of signal can be written as
\begin{equation}\label{A-3-3}
\begin{array}{l}
\rho _{s_h }  = Tr_i (\left| {\psi _{SPDC_h } } \right\rangle \left\langle {\psi _{SPDC_h } } \right|) = \sum_{n = 1}^\infty  {Pr_{n - 1} } \left| n \right\rangle _s \left\langle n \right| \\
= \sum_{n = 1}^\infty  ( 1 - \gamma ^2 )\gamma ^{2n - 2} \left| n \right\rangle _s \left\langle n \right|. \\
\end{array}
\end{equation}
In this case, the $g^{(2)}(0)$ can be obtained as
\begin{equation}\label{A-3-4}
g^{(2)} _{s_h } (0) = \frac{{\sum_{n = 1}^\infty  {n(n - 1)} p_{n - 1} }}{{(\sum_{n = 1}^\infty  n p_{n - 1} )^2 }}.
\end{equation}
At low pump power, $g^{(2)}_{s_h } (0)$ is near 0.
\begin{equation}\label{A-3-5}
g^{(2)} _{s_h } (0)  \to \frac{{2Pr_1 }}{{(Pr_0 )^2 }} = \frac{{2\gamma ^{2} }}{{(1 - \gamma ^2 )}} \to 0.
\end{equation}

\section{ $g^{(2)}(0)$ measured in experiment: $g_{exp}^{(2)} (0)$ }
In this section, we review the experimental  definition of $g^{(2)}(0)$  by U'Ren et al \cite{URen2005a} and Mosley \cite{MosleyPhD}.
With the setup in Fig.\,\ref{setup}(b),
the probability of detecting a signal photon in either SNSPD2 or SNSPD3 for a heralding event in the idler is
\begin{equation}\label{A-4-1}
P_{h_1}  = \frac{{CC_{12}  + CC_{13} }}{{SC_1 }},
\end{equation}
and the probability of detecting two signal photons for a heralding event in the idler is
\begin{equation}\label{A-4-2}
P_{h_2}  = \frac{{CC_{123} }}{{SC_1 }}.
\end{equation}
The value of $g^{(2)}(0)$ is obtained from the ratio of generating two signal photons for one heralding event to the square of the probability of generating only one
\begin{equation}\label{A-4-3}
g_{exp}^{(2)} (0) = \frac{{2P_{h_2} }}{{(  P_{h_1} )^2 }} = \frac{{2SC_1  \times CC_{123} }}{{(CC_{12}  + CC_{13} )^2 }},
\end{equation}
where the factor of 2 is added in the numerator because half the coincidence is lost after a 50/50 beam splitter.
From this deduction process,  we can learn that this experimental definition  of  $g^{(2)}(0)$  \cite{URen2005a, MosleyPhD} is only  valid for  the case where two-pair components are dominant.
Next, we consider  all components in SPDC, which is important for high-pump power SPDC.

\section{  $g^{(2)}(0)$ expected in experiment:  $g_{exp2}^{(2)} (0)$}

In this section, we calculate  $g^{(2)}(0)$ following  Eq.(\ref{A-4-3}), but with the parameters of $p$, $\tau $  and $\eta_{1(2)}$  in Tab. \ref{tableCC}.
With the setup in Fig.\,\ref{setup}(b), when one pulse of the signal with N photons, i.e., $\left| n \right\rangle$ state, is sent to a FBS, the output state  is
\begin{equation}\label{A-5-1}
\frac{1}{{\sqrt {n!} }}\frac{1}{{\sqrt {2^n } }}\sum_{k = 0}^n {\frac{{n!}}{{\sqrt {(n - k)!} \sqrt {k!} }}} \left| {n - k} \right\rangle \left| k \right\rangle.
\end{equation}
The output state is detected by SNSPD2 and SNSPD3 with a coincidence count of
\begin{equation}\label{A-5-2}
\frac{1}{{2^n n!}}\sum_{k = 1}^{n - 1} {\frac{{(n!)^2 }}{{(n - k)!k!}}} (1 - (1 - \eta_2 )^{n - k} )(1 - (1 - \eta_3 )^k ).
\end{equation}
Therefore the three-fold coincidence counts between SNSPD1-3 are
\begin{equation}\label{A-5-3}
\begin{array}{l}
  CC_{123}  = f \times  \sum\limits_{n = 1}^\infty  {\{ (1 - p\tau )(p\tau )^{\rm{n}} (} 1 - (1 - \eta _1 )^n )\frac{1}{{2^n n!}} \\
 \times   \sum\limits_{k = 1}^{n - 1} {\frac{{(n!)^2 }}{{(n - k)!k!}}} (1 - (1 - \eta _2 )^{n - k} )(1 - (1 - \eta _3 )^k )\}.  \\
 \end{array}
\end{equation}
The coincidence counts between the idler and the two arms of the signal are
\begin{equation}\label{A-5-4}
\begin{array}{l}
  CC_{12}  = f \times \sum\limits_{n = 1}^\infty  {\{ (} 1 - (1 - \eta _1 )^n )(1 - p\tau )(p\tau )^n \frac{1}{{n!2^n }}  \\
 \times   \sum\limits_{k = 0}^{n - 1} {\frac{{(n!)^2 }}{{(n - k)!k!}}} (1 - (1 - \eta _2 )^{n - k} )\}. \\
 \end{array}
 \end{equation}
 and
 \begin{equation}\label{A-5-5}
 \begin{array}{l}
 CC_{13}  = f \times \sum\limits_{n = 1}^\infty  {\{ (} 1 - (1 - \eta _1 )^n )(1 - p\tau )(p\tau )^n \frac{1}{{n!2^n }}  \\
 \times   \sum\limits_{k = 1}^n {\frac{{(n!)^2 }}{{(n - k)!k!}}} (1 - (1 - \eta _3 )^k )\}.\\
  \end{array}
 \end{equation}
The single count  of the idler is
\begin{equation}\label{A-5-6}
SC_1  = f \times \sum_{n = 1}^\infty ( 1 - (1 - \eta )^n )(1 - p\tau )(p\tau )^{n}.
\end{equation}
Therefore, the $g^{(2)}(0)$ is measured as
\begin{widetext}
\begin{equation}\label{A-5-7}
\begin{array}{l}
  g_{exp2}^{(2)} (0) = \frac{{2SC_1  \times CC_{123} }}{{(CC_{12}  + CC_{13} )^2 }}  =
 \frac{{2\sum\limits_{n = 1}^\infty  ( 1 - (1 - \eta _1 )^n )(1 - p\tau )(p\tau )^n  \times \sum\limits_{n = 1}^\infty  {\{ (1 - p\tau )(p\tau )^n (} 1 - (1 - \eta _1 )^n )\frac{1}{{2^n n!}}\sum\limits_{k = 1}^{n - 1} {\frac{{(n!)^2 }}{{(n - k)!k!}}} (1 - (1 - \eta _2 )^{n - k} )(1 - (1 - \eta _3 )^k )\} }}{{\{ \sum\limits_{n = 1}^\infty  ( 1 - (1 - \eta _1 )^n )(1 - p\tau )(p\tau )^n \frac{1}{{n!2^n }}[\sum\limits_{k = 0}^{n - 1} {\frac{{(n!)^2 }}{{(n - k)!k!}}} (1 - (1 - \eta _2 )^{n - k} ) + \sum\limits_{k = 1}^n {\frac{{(n!)^2 }}{{(n - k)!k!}}} (1 - (1 - \eta _3 )^k )]\} ^2 }}.
 \end{array}
\end{equation}
\end{widetext}

\section{ $g^{(3)}(0)$ for both the  signal and idler (squeezed state: $g_{s,i}^{(3)} (0)$), and the unheralded signal (thermal state: $g_s^{(3)} (0)$) }
The definition of $g^{(3)}(0)$ is
\begin{equation}\label{A-6-1}
 g_{}^{(3)} (0) \equiv \frac{{\left\langle {\hat a^\dag  \hat a^\dag  \hat a^\dag  \hat a\hat a\hat a} \right\rangle }}{{\left\langle {\hat a^\dag  \hat a} \right\rangle ^3 }} = \frac{{\left\langle {\hat n^3 } \right\rangle  - 3\left\langle {\hat n^2 } \right\rangle  + 2\left\langle {\hat n} \right\rangle }}{{\left\langle {\hat n} \right\rangle ^3 }}.
\end{equation}
$g^{(3)}(0)$ for both the signal and idler (squeezed state) is calculated as
\begin{equation}\label{A-6-2}
\begin{array}{l}
g_{s,i}^{(3)} (0) = \frac{{\sum\limits_{n = 0}^\infty  {((2n)^3  - 3(2n)^2  + 2(2n))} Pr_n }}{{(\sum\limits_{n = 0}^\infty  {(2n)} Pr_n )^3 }} \to \frac{{2Pr_2 }}{{(2Pr_1  + 4Pr_2 )^3 }}.
\end{array}
\end{equation}
$g^{(3)}(0)$ for the unheralded signal (thermal state) is calculated as
\begin{equation}\label{A-6-3}
\begin{array}{l}
g_s^{(3)} (0) = \frac{{\sum\limits_{n = 0}^\infty  {(n^3  - 3n^2  + 2n)} Pr_n }}{{(\sum\limits_{n = 0}^\infty  n Pr_n )^3 }} \to \frac{{6Pr_3 }}{{(Pr_1  + 2Pr_2  + 3Pr_3 )^3 }}.
\end{array}
\end{equation}


\end{document}